\documentclass[seceq,preprint]{ptptex}

\usepackage{bm}
\usepackage{latexsym}
\usepackage{amsfonts}
\usepackage{amssymb}
\usepackage{amsmath}
\def\square{\mathchoice\sqr54\sqr54\sqr{2.1}3\sqr{1.5}3} 
\def\sqr#1#2{{\vcenter{\vbox{\hrule height.#2pt\hbox{\vrule
width.#2pt height#1pt \kern#1pt\vrule width.#2pt}\hrule height.#2pt}}}}
\def\square{\mathchoice\sqr54\sqr54\sqr{2.1}3\sqr{1.5}3} 

\begin{document}

\preprintnumber[3cm]{
YITP-07-76\\
November 7, 2007}

\markboth{
N.~Deruelle, M.~Sasaki and Y.~Sendouda %
}{
Junction Conditions in $f(R)$ Theories of Gravity
}
\title{Junction Conditions in $\bm{f(R)}$ Theories of Gravity}

\author{Nathalie Deruelle$^1$, Misao Sasaki$^2$ and Yuuiti Sendouda$^2$}

\inst{$^1$APC,UMR 7164
 (CNRS, Universit\'e Paris 7, CEA, Observatoire de Paris),\\
10, rue Alice Domon et Lonie Duquet,
75205 Paris Cedex 13\\
$^2$Yukawa Institute for Theoretical Physics, Kyoto University,\\
Kyoto 606-8502, Japan}

%
%
%

\abst{
Taking advantage of the conformal equivalence of $f(R)$ theories 
of gravity with General Relativity coupled to a scalar field
 we generalize the Israel junction conditions for this class of
theories by direct integration of the field equations. 
We suggest a specific non-minimal coupling of matter to gravity 
which opens the possibility of a new class of braneworld scenarios.
}

\maketitle

\section{Introduction} 
\label{introduction}

In 1918 Weyl \cite{Weyl:1918ibWeyl:1919fiWeyl:1921} was the first to consider Lagrangians for gravity
which are a linear combination of the three quadratic scalars
$R^2$, $R_{AB}R^{AB}$ and $R_{ABCD}R^{ABCD}$ formed out of the scalar curvature,
the Ricci and Riemann tensors of a metric 
$g_{AB}$\footnote{The fact that the $R_{ABCD}R^{ABCD}$ can be traded
for $-(R^2-4R_{AB}R^{AB})$ (in $4$ dimensions) was apparently first shown by Bach,
and then by Lanczos \cite{Bach:1921Lanczos:1938sf} (the Gauss--Bonnet theorem \cite{Chern:1944}).
For a historical review, see \citen{Schmidt:2006jt}.}.

Pauli \cite{Pauli:1919} and Eddington \cite{Eddington:1924} then noticed that 
the Schwarzschild metric was a solution of the corresponding vacuum
field equations. Since at that time all precision tests of General Relativity
theories relied on the Schwarzschild metric, these authors concluded 
that quadratic theories were a priori just as viable as ordinary 
General Relativity based on the Einstein--Hilbert Lagrangian $R$.
(As noted by Chiba {\it et al}., \cite{Chiba:2006jp} this seems to be still 
a fairly widespread belief.)

The question of the unicity of the Schwarzschild metric in such theories 
in vacuo was first investigated by Buchdahl \cite{Buchdahl:1961} who showed that
the Schwarzschild metric was the only spherically symmetric, asymptotically
flat, solution of the vacuum pure $R^2$ field equations. The subject was 
considered anew much later by Whitt \cite{Whitt:1984pd}
 (in the case of the $R+a_2R^2$ Lagrangian in $4$ dimensions),
and Mignemi and Wiltshire \cite{Mignemi:1991wa} who showed 
that the Schwarzschild metric was the only $D$-dimensional,
spherically symmetric, asymptotically flat, vacuum solution 
with a regular horizon, for all polynomial $f(R)=R+\sum a_nR^n$ 
with $a_2>0$\footnote{As emphasized in \citen{Mignemi:1991wa}, 
other solutions exist if one relaxes the condition of asymptotic 
flatness and allows for asymptotically de Sitter or anti-de Sitter spacetimes.}.

Now, since the $f(R)$ field equations are fourth order differential
equations for the metric, they possess ``runaway'' solutions on top of 
solutions which smoothly tend to solutions of the Einstein equations in 
the limit $f(R)\to R$\footnote{This fact was used by
 Starobinsky \cite{Starobinsky:1980te} to build the first inflationary cosmological model.}.
The question of whether the Schwarzschild metric is the $f(R)$ solution 
outside a distribution of ordinary matter  (rather than a black hole),
either point-like or extended,  must therefore be raised.
Pechlaner and Sexl \cite{Pechlaner:1966} showed that, in fact,
 in pure $R^2$ theory the metric cannot be asymptotically flat as soon as
the field equations have a right-hand side describing matter with positive
energy density. They also showed that in $R+a_2R^2$ theory the metric can be
asymptotically flat but that, at linear order around Minkowski spacetime,
it is not the linearized Schwarzschild metric.
The origin of such results is clearly explained by
Havas \cite{Havas:1977}\,: The Green function for the (second order) Einstein
equations, which at lowest order reduce to $\nabla^2G_N=-\delta(r)$,
takes the familiar form $G_N=1/r$, which yields the linearized Schwarzschild 
solution. On the other hand, in the case of the (fourth order) 
pure $R^2$ theory the Green function solving the linearized field 
equations $\nabla^4G=-\delta(r)$ is $G=r/2$,  which yields 
a divergent metric. As for  $G_N=1/r$ it satisfies 
$\nabla^4 G_N=-\nabla^2\delta(r)$\,; this means that in
 pure $R^2$ theory the source for the Schwarzschild metric 
is not a delta function but its second derivative which does not 
represent a point-like distribution of matter with positive energy density. 
Finally the Green function for $R+a^2R^2/6$ theory is $G=(1-\mathrm e^{-r/a})/r$,
yielding the Pechlaner--Sexl metric (after correcting a couple of typos
in their equation (23)).
(See also \citen{Stelle:1977ry,Eddington:1924}.)

This question of unicity or non-unicity of the Schwarzschild
 (or Schwarzschild--(A)dS) solution was revived recently when
$f(R)$ theories were invoked in an attempt to explain the observed
 present acceleration of the universe by means other than a 
cosmological constant \cite{Capozziello:2003tkCarroll:2003wy} (see also \citen{Nojiri:2007uq} and references therein).
 In that context the question became whether or not the Schwarzschild
(or Schwarzschild--de Sitter) metric is, at least approximately,
a solution of the $f(R)$ field equations in the presence of
localized sources such as the Sun. 
In \citen{Henttunen:2007bz}, static spherically symmetric solutions for
the special case $f(R)=R-\mu^4/R$ were built numerically with
matter represented by a perfect fluid (see also references in \citen{Henttunen:2007bz}).
It was found that, if the metric tends to an appropriate de Sitter
limit to explain the acceleration of the universe then the PPN
parameter $\gamma$ measuring, e.g., the light bending by the Sun is of
order $1/2$ instead of $1$ \cite{Chiba:2003ir},
which rules out these ``dark energy'' models.
In \citen{Chiba:2006jp} and \citen{Faulkner:2006ub} (see also references therein), the field equations were solved 
at the linear approximation around Minkowski spacetime and 
similar results were found for a wide class of $f(R)$ theories.

In many of the above mentioned analyses advantage is taken from the fact
that $f(R)$ field equations  are conformally equivalent to Einstein
gravity with a minimally coupled scalar field (this was first shown by
Higgs \cite{Higgs:1959} for $f(R)=R^2$ and by Teyssandier and Tourrenc \cite{Teyssandier:1983zz}
in the general case).
This property puts the theories on a more familiar footing but,
in itself, does not modify either the mathematics or the physics of 
the problem. 
Nevertheless conclusions concerning what are the ``correct''
solutions of the field equations are still controversial.
In particular the validity
of the linear approximation has been challenged (see e.g.\ \citen{Navarro:2006mw} and references therein) and
the question of how one recovers the (de Sitter--) Schwarzschild solution 
in the Einstein limit does not seem to be settled yet (see e.g.\ \citen{Olmo:2006eh} and references therein).

In this paper, taking advantage of this conformal equivalence, 
we investigate junction conditions for a brane
(i.e. an infinitesimally thin domain wall) in $f(R)$ theories.
The topic has already been investigated in \citen{Parry:2005ebNojiri:2001aeNojiri:2000gv}, where, however, 
the questions of the Einstein limit and the coupling to matter on the brane
were not addressed. Here, as shown below, we are
able to see clearly the possible irregularity which may appear
in the Einstein limit $f(R)\to R$. Then we suggest a specific
non-minimal coupling of matter on the brane to gravity 
which opens the possibility of a new class of braneworld scenarios.

The paper is organized as follows. In Sec.~\ref{weakjunction},
we consider general quadratic theories of gravity and
discuss possible freedom in the choice of
the junction conditions depending on how singular one
allows the metric to be. Then, as an example, we formulate the 
junction conditions by requiring the metric to be least singular,
namely its first and second derivatives are continuous across
the brane.
Then, in Sec.~\ref{Jordanframe}, focusing of $f(R)$ theories, 
we formulate the junction conditions which allow for discontinuities
in the first derivatives of the metric.
In Sec.~\ref{Einsteinframe}, we consider $f(R)$ theories in
the Einstein frame, i.e., Einstein--scalar theories conformally
equivalent to $f(R)$ theories, and formulate the junction conditions
in the Einstein frame. In doing so, we suggest a new type of
gravitational coupling for the matter on the brane.
Finally,  in Sec.~\ref{Jordanframebis}, we translate these generalized
 junction conditions back to the original, Jordan, frame.
We work in $D$ spacetime dimensions. The gravitational constant
$8\pi G$ in $D$ dimensions is set to unity; the signature is $-++\cdots+$.

\section{``Weak'' junction conditions in quadratic theories of gravity}
\label{weakjunction}

Consider the quadratic Lagrangian,
\begin{equation}
L=-2\Lambda+R+\gamma\,R_{ABCD}R^{ABCD}-4\beta \,R_{AB}R^{AB}+\alpha\, R^2\,.
\end{equation}
The variational derivative of $\sqrt{-g} L$
with respect to the metric $g_{AB}$ yields, up to a divergence,
$\delta (\sqrt{-g}\,L)=-\sqrt{-g}\, \sigma^{AB}\,\delta g_{AB}$
with (see e.g.\ \citen{Deruelle:2003ck})
\begin{equation}
\begin{aligned}
\sigma_{AB}
&= -\frac{1}{2}\,L\, g_{AB}+R_{AB}
\\
&\quad +2\gamma\, R_{ALNP}\,R_B{}^{LNP}+4(2\beta-\gamma)R^C{}_D\,R^D{}_{BAC}
-4\gamma\, R_A{}^C\,R_{CB}+2\alpha \, R\, R_{AB}
\\
&\quad +4(\gamma-\beta)\square R_{AB}
+2(\alpha -\beta)g_{AB}\square\, R-2(\alpha +\gamma-2\beta)D_{A}D_{B}\,R
\end{aligned}
\end{equation}
where $D$ is the covariant derivative associated with $g_{AB}$.
The field equations are, in the bulk,
\begin{equation}
\sigma_{AB}=0\,.
\end{equation}
These equations are fourth order in the derivatives of the metric, 
except for the Gauss--Bonnet combination
$\alpha=\beta=\gamma$ that we shall exclude\footnote{The junction conditions in 
Einstein--Gauss--Bonnet theory are well-known.
See \citen{Davis:2002gnGravanis:2002wy}, and, e.g.\ \citen{Deruelle:2000ge}.}.

We use Gaussian-normal coordinates in which the metric is 
\begin{equation}
\mathrm ds^2=\mathrm dy^2+\gamma_{\mu\nu}\mathrm dx^\mu\mathrm dx^\nu\,,
\end{equation}
where the brane is assumed to be located at $y=0$.
In terms of the extrinsic curvature,
\begin{equation}
K_{\mu\nu}=-\frac{1}{2}\frac{\partial\gamma_{\mu\nu}}{\partial y}\,,
\end{equation}
 the Riemann tensor is
\begin{equation}
\begin{aligned}
R^y_{\ \mu y\nu}
&=\frac{\partial K_{\mu\nu}}{\partial y}+K_{\rho\nu}K^\rho{}_\mu\,,
\quad
R_{y\mu\nu\rho}=\bar D_\nu K_{\mu\rho}-\bar D_\rho K_{\mu\nu}\,,
\\
R_{\lambda\mu\nu\rho}
&=\bar R_{\lambda\mu\nu\rho}+K_{\mu\nu}K_{\lambda\rho}-K_{\mu\rho}K_{\lambda\nu}\,,
\end{aligned}
\end{equation}
where $\bar D_\rho$ and $\bar R^\mu_{\ \nu\rho\sigma}$ are the 
covariant derivative and the Riemann tensor associated with the metric
$\gamma_{\mu\nu}$, with the Greek indices
being raised or lowered by the metric $\gamma_{\mu\nu}$.

For convenience, we introduce the tensor
\begin{equation}
H_{\mu\nu}\equiv\frac{\partial^2K_{\mu\nu}}{\partial y^2}
=-\frac{1}{2}\frac{\partial^3\gamma_{\mu\nu}}{\partial y^3}\,.
\end{equation}
Keeping in $\sigma_{AB}$ only the terms with highest derivatives in $y$,
i.e. the terms proportional to $\partial H_{\mu\nu}/\partial y$,
one finds (see e.g.\ \citen{Deruelle:2003ck}) that there are no such terms
in $\sigma_{yy}$ and $\sigma_{y\mu}$,
and that they appear in $\sigma_{\mu\nu}$ under the following combination\,:
\begin{equation}
\sigma_{\mu\nu}=4\frac{\partial}{\partial y}
[(\gamma-\beta)H_{\mu\nu}+(\alpha-\beta)\gamma_{\mu\nu}H]+\cdots,
\end{equation}
with $H=\gamma^{\rho\sigma}H_{\rho\sigma}$. 

Suppose now that there exists a sub-class of metrics 
$\gamma_{\mu\nu}(y,x^\rho)$ solving the bulk equations $\sigma_{AB}=0$,
whose third order derivatives $H_{\mu\nu}$ jump across $y=0$; 
that is, such that $H_{\mu\nu}$ can be written as, e.g., 
$H_{\mu\nu}=h_{\mu\nu}(x^\rho)\tanh(y/\ell)$ with $\ell\to0$,
in a region of order $\ell$ around $y=0$ (``$\mathbb Z_2$-symmetric'' case). 
For this sub-class of metrics $\sigma_{\mu\nu}$ exhibits a Dirac 
distribution-like behavior at $y=0$\,:
\begin{equation}
4\frac{\partial}{\partial y}[(\gamma-\beta)H_{\mu\nu}
+(\alpha-\beta)\gamma_{\mu\nu}H]\equiv\delta(y)D_{\mu\nu}\,,
\end{equation}
where $D_{\mu\nu}$ is the ``strength'' of the singularity.
Integration across the brane then yields
\begin{equation}
4[(\gamma-\beta)H_{\mu\nu}+(\alpha-\beta)\gamma_{\mu\nu}H]^+_-=D_{\mu\nu}\,,
\label{jumpI}
\end{equation}
where $[F]^+_-\equiv \lim_{y\to0+}F(y)-\lim_{y\to0-}F(y)$\,. 
If we require this class of metrics to be solutions of the
field equations, then we must have 
\begin{equation}
D_{\mu\nu}=S_{\mu\nu}\,,
\label{conditionI}
\end{equation}
where we may naturally interpret $S_{\mu\nu}$ as the total
energy--momentum tensor of the brane.
These equations, together with Eq.~(\ref{jumpI}),
give the junction conditions.

These junction conditions were first given in \citen{VonBorzeszkowski:1980ka} (in the
particular case $D=4$ and hence $\gamma=0$, because of the
Gauss--Bonnet theorem) and generalized recently in \citen{Balcerzak:2007da} 
to Lagrangians of the type 
$L=f(R, R_{AB}R^{AB},R_{ABCD}R^{ABCD})$. It is clear that such 
discontinuities are specific to higher derivative theories and smoothly 
disappear in the Einstein limit when the parameters 
$\alpha$, $\beta$ and $\gamma$ are ``switched off.''

An important point (apparently overlooked in \citen{Balcerzak:2007da}) is that
the contracted Bianchi identities, $D_B\sigma^B{}_A\equiv0$, imply
that the brane energy--momentum tensor is conserved\,:
\begin{equation}
\bar D_\nu S^\nu{}_\mu=0\,.
\end{equation}

We leave to further work the question of finding a bulk metric, 
a solution to the bulk equations (e.g.\ anti-de Sitter), which can be 
written in Gaussian coordinates in such a way as to exhibit a 
discontinuity in its third derivative across $y=0$. 
(The forms of the metric given in, e.g., \citen{Binetruy:1999hyMukohyama:1999qx} do not 
belong to this desired sub-class as they exhibit a jump in
their {\it first} derivatives.)

A few remarks to conclude this section are in order\,:
\begin{list}{}{}
\item[(a)]
For $\gamma=2\beta-\alpha$ (generalization of Eddington's 
choice \cite{Eddington:1953}),
the junction conditions (\ref{conditionI}) take the form
(in the $\mathbb Z_2$-symmetric case)
\begin{equation}
H_{\mu\nu}-\gamma_{\mu\nu} H=\frac{S_{\mu\nu}}{8 (\beta-\alpha)}\,.
\end{equation}

\item[(b)]
For $\alpha-\beta=-\frac{\gamma-\beta}{D-1}$ (that is Weyl's choice
in $D=4$ and with $\gamma=0$ \cite{Weyl:1918ibWeyl:1919fiWeyl:1921}), they become
\begin{equation}
H_{\mu\nu}-\frac{H}{D-1}\gamma_{\mu\nu}=\frac{S_{\mu\nu}}{8 (\gamma-\beta)}\,.
\end{equation}
Therefore the total energy--momentum tensor on the brane must be traceless.

\item[(c)]
Finally in the pure $f(R)$ case ($\beta=\gamma=0$, $L=-2\Lambda+R+\alpha R^2$),
the total
energy--momentum tensor on a $\mathbb Z_2$-symmetric brane is constrained to be of the form
\begin{equation}
S_{\mu\nu}=8\alpha H\gamma_{\mu\nu}\,.
\label{caseII}
\end{equation}
The fact that $S_{\mu\nu}$ is conserved implies $H$ is a constant.
Thus, the matter on the brane must be vacuum energy.
\end{list}

This final example indicates that imposing the metric to be of class
$C^2$ (continuity of $\gamma_{\mu\nu}$ and its first and second derivatives)
is probably too restrictive to allow for physically interesting braneworld
scenarios in higher derivative theories of gravity.
We shall therefore seek junction conditions that allow 
for more singular metrics.
In the rest of the paper, we concentrate on $f(R)$ theories.

\section{Junction conditions in the Jordan frame\,:
 the standard approach}
\label{Jordanframe}

The field equations derived from the variation of $\sqrt{-g}f(R)$ 
with respect to the metric $g_{AB}$
are $\sigma_{AB}=0$ in the bulk, with 
\begin{equation}
\sigma_{AB}=f'(R)G_{AB}
+\frac{1}{2}g_{AB}(R\,f'(R)-f(R))+g_{AB}\square f'(R)-D_{A}D_{B}f'(R)\,,
\label{Jfeq}
\end{equation}
where $f'(R)\equiv \mathrm df/\mathrm dR$ and $G_{AB}=R_{AB}-\frac{1}{2}g_{AB}R$ 
is the Einstein tensor\footnote{We consider here metric $f(R)$ theories.
For variation a la Palatini, see \citen{Ferraris:1982Sotiriou:2006qn} and references therein.}.
In a Gaussian normal coordinate system, 
$\mathrm ds^2=\mathrm dy^2+\gamma_{\mu\nu}\mathrm dx^\mu\mathrm dx^\nu$,
the Einstein tensor is decomposed as
\begin{equation}
\begin{aligned}
G_{yy}&=-\frac{1}{2}(K_{\mu\nu}K^{\mu\nu}-K^2+\bar R)\,,
\\
G_{y\mu}&=-\bar D_\nu (K^\nu{}_\mu-\delta^\nu_\mu K)\,,
\\
G_{\mu\nu}&=\partial_y(K_{\mu\nu}-K\gamma_{\mu\nu})
+2K^\rho{}_\mu K_{\rho\nu}-3KK_{\mu\nu}
+\frac{1}{2}\gamma_{\mu\nu}(K_{\alpha\beta}K^{\alpha\beta}+K^2)
+\bar G_{\mu\nu}\,,
\end{aligned}
\label{Gdecomp}
\end{equation}
where the extrinsic curvature is
 $K_{\mu\nu}\equiv -\frac{1}{2}\partial_y\gamma_{\mu\nu}$. We note that
\begin{equation}
\begin{aligned}
R&=2\partial_yK-K_{\mu\nu}K^{\mu\nu}-K^2+\bar R
\\
&=2\partial_yK-K^*_{\mu\nu}K^{*\mu\nu}-\frac{D}{D-1}K^2+\bar R\,,
\end{aligned}
\label{Rdecomp}
\end{equation}
where $K^*_{\mu\nu}$ is the traceless part of the extrinsic
curvature\,: $K^*_{\mu\nu}\equiv K_{\mu\nu}-\frac{K}{D-1}\gamma_{\mu\nu}$.

For convenience, we decompose $\sigma_{AB}$ into 
\begin{equation}
\sigma_{AB}=Q_{AB}+L_{AB}\,,
\label{sigmadecomp}
\end{equation}
with
\begin{equation}
\begin{aligned}
Q_{AB}&=f'(R)\,G_{AB}+\frac{1}{2}(R\,f'(R)-f(R))\,g_{AB}\,,
\\
L_{AB}&=-D_{A}D_{B}f'(R)+g_{AB}\square f'(R)\,.
\end{aligned}
\end{equation}
Their components are given by
\begin{equation}
\begin{aligned}
Q_{yy}&=\frac{1}{2}(R\,f'(R)-f(R))+f'(R)\,G_{yy}\,,
\\
Q_{y\mu}&=f'(R)G_{y\mu}\,,
\\
Q_{\mu\nu}&=\frac{1}{2}\gamma_{\mu\nu}(R\,f'(R)-f(R))+f'(R)G_{\mu\nu}\,,
\end{aligned}
\end{equation}
and
\begin{equation}
\begin{aligned}
L_{yy}&=-K\partial_yf'(R)+\bar{\square}f'(R)\,,
\\
L_{y\mu}&=-\partial_\mu\partial_y f'(R)-K^\nu_\mu\partial_\nu f'(R)\,,
\\
L_{\mu\nu}&=
-\bar D_{\mu\nu}f'(R)+K_{\mu\nu}\partial_yf'(R)
+\gamma_{\mu\nu}(\partial_{yy}f'(R)+\bar{\square}f'(R)-K\partial_yf'(R))\,.
\end{aligned}
\end{equation}

In Sec.~\ref{weakjunction} we considered the class of metrics $\gamma_{\mu\nu}$
which were continuous across $y=0$ with  continuous  first and second 
derivatives. In that case all the components of the Einstein tensor and
of $Q_{AB}$ are well behaved and a Dirac distribution  appears in
$L_{\mu\nu}$ in the term $\gamma_{\mu\nu}\partial_{yy}f'(R)$. 
Now, since, for that sub-class, $R$ is continuous and hence $\partial_yR$ 
is at most discontinuous, the delta distribution behavior of 
$\partial_{yy}f'(R)$ comes from $\partial_{yy}R$\,:
\begin{equation}
\begin{aligned}
\partial_{yy}f'(R)
&=f''(R)\partial_{yy}R+f'''(R)(\partial_yR)^2
\\
&=2f''(R)\partial_{yyy}K+\cdots
=2f''(R)\partial_yH+\cdots\,,
\end{aligned}
\label{dyyfp}
\end{equation}
with $H_{\mu\nu}=\partial_{yy}K_{\mu\nu}$.
In the particular case $f(R)=-2\Lambda+R+\alpha R^2$
we thus  recover the result (\ref{caseII}) of Sec.~\ref{weakjunction}.

Here, on the other hand, we shall consider the class of metrics 
which are continuous across $y=0$ but which allow for (a certain type
of) discontinuity in their first derivatives.

The scalar curvature $R$ which could be now, a priori, a delta function,
must be at most discontinuous\,; otherwise unacceptable $(\delta(y))^2$ terms
would appear in $Q_{AB}$ (unless of course $f(R)=-2\Lambda+R$). 
Furthermore, an inspection of $L_{\mu\nu}$ tells us that 
$R$ must be in fact continuous across $y=0$\,; otherwise a $(\delta(y))^2$
term would appear in $L_{\mu\nu}$
(unless $f(R)$ is quadratic, because it precisely arises from
the term $f'''(R)(\partial_yR)^2$, see (\ref{dyyfp})).
We shall therefore restrict ourselves to the sub-class of metrics
with continuous scalar curvature $R$.


Again we leave to further work the question of finding a bulk metric, 
solution to the bulk equations, which can be written 
in Gaussian coordinates  in such a way as to exhibit  discontinuities in 
its first order derivative across $y=0$, but not in $R$. (It seems that
the fact that the scalar curvature must be continuous has been
overlooked in the recent paper \cite{Afonso:2007gc}.)

We then see by inspection that the $yy$ and $y\mu$ components of
$\sigma_{AB}$ at most jump across $y=0$ and that the delta-like part
of $\sigma_{\mu\nu}$ is
\begin{equation}
\partial_y[f'(R)(K_{\mu\nu}-K\gamma_{\mu\nu})
+\gamma_{\mu\nu}f''(R)\partial_yR]\equiv\delta(y)D_{\mu\nu}\,.
\label{fsingular}
\end{equation}
Integration across the brane then yields the junction conditions
\begin{equation}
D_{\mu\nu}
=[f'(R)(K_{\mu\nu}-K\gamma_{\mu\nu})+\gamma_{\mu\nu}f''(R)\partial_yR]^+_-
=S_{\mu\nu}\,,
\label{JfKjunction}
\end{equation}
where $S_{\mu\nu}$ is the total energy momentum tensor on the brane.
From the contracted Bianchi identities,
we have that it must be conserved\,: $\bar D_\nu S^\nu{}_\mu=0$.

Note that when $f(R)\to R$ the junction conditions (\ref{JfKjunction})
do not reduce to the familiar Israel conditions \cite{Lanczos} 
as they have to be supplemented by the condition of continuity of $R$.
What happens when $f(R)=-2\Lambda+R+\ell^2 R^2+...$ when $\ell \to0$ is
that the bulk geometry may approach a solution of the Einstein 
bulk equations (e.g., AdS) everywhere, to the exception of
a region of size $\ell$ in the vicinity of the brane, so that
when $\ell$ becomes very small the thin shell limit is no longer
valid and the thickness of the brane must be taken into account.
To render this irreducible difference between Einstein and $f(R)$
theories manifest, let us split the junction conditions into 
their trace and traceless parts, recalling 
that $\gamma_{\mu\nu}$ as well as $R$ are continuous (see (\ref{Rdecomp}))\,:
\begin{equation}
\begin{aligned}
&[\gamma_{\mu\nu}]^+_-=0\,,
\\
&[R]^+_-=0\qquad\Longrightarrow\qquad[K]^+_-=0
\,,\quad[2\partial_yK-K^*_{\mu\nu}K^{*\mu\nu}]^+_-=0\,,
\\
&f'(R)[K^*_{\mu\nu}]^+_- = S^*_{\mu\nu}\,,
\\
&(D-1)f''(R)[\partial_{y}R]^+_-= S\,,
\end{aligned}
\label{Jfjunction}
\end{equation}
where $K^*_{\mu\nu}$ and $S^*_{\mu\nu}$ are the traceless parts of 
the extrinsic curvature and brane energy--momentum tensor, respectively. 
The ``weak'' junction conditions considered in Sec.~\ref{weakjunction} 
are just a particular sub-class of the above such that $S^*_{\mu\nu}=0$.

For further reference, we shall also generalize them to a non-Gaussian
coordinate system, $\mathrm ds^2=N^2\mathrm dy^2+\gamma_{\mu\nu}\mathrm dx^\mu\mathrm dx^\nu$,
where $N$ is a continuous lapse function\,:
\begin{equation}
\begin{aligned}
&[\gamma_{\mu\nu}]^+_-=0\,,
\quad[K]^+_-=0\,,
\quad[R]^+_-=0\,,
\\
&f'(R)[K^*_{\mu\nu}]^+_- = S^*_{\mu\nu}\,,
\\
&(D-1)f''(R)\frac{1}{N}[\partial_{y}R]^+_-= S\,,
\end{aligned}
\label{Jfjunctionbis}
\end{equation}
where, now, the extrinsic curvature is defined as
  $K_{\mu\nu}=-\frac{1}{2N}\partial_y\gamma_{\mu\nu}$.

In this and the preceding section we just stated that the jumps 
in some derivatives of the bulk metric coefficients must be equal
to the {\it total} energy--momentum tensor of matter on the brane. 
We have yet to decide on the type of matter we want to have on the brane.
An even more crucial issue is to decide how matter on the brane couples
to gravity, that is, how it couples to the metric $\gamma_{\mu\nu}$ 
(which is the only dynamical variable at our disposal). 

It is natural (and this is the hypothesis which is ``universally'' made in 
the literature) to assume that matter is minimally coupled to the metric. 
For example, if matter on the brane is taken to be a scalar field $\psi$
 with potential $V(\psi)$, we have
\begin{equation}
S_{\mu\nu}=\partial_\mu\psi\partial_\nu\psi
-\gamma_{\mu\nu}
\left(\frac{1}{2}\gamma^{\alpha\beta}\partial_\alpha\psi\partial_\beta\psi
+V(\psi)\right)\,.
\label{Jmatter}
\end{equation}
As we shall see in an accompanying paper \cite{paper2},
the junction conditions (\ref{Jfjunction}) and (\ref{Jfjunctionbis}) thus obtained are the standard ones, that is,
those one derives from the action $\frac{1}{2}\int\!\mathrm d^Dx\,\sqrt{-g}f(R)$
supplemented by the Hawking--Luttrell
 boundary term $\int\!\mathrm d^{D-1}x\sqrt{-\gamma}f'(R)K$ \cite{Hawking:1984ph}.

In the next section, we shall question this conventional wisdom, 
taking advantage of the conformal equivalence of $f(R)$ theories of 
gravity with General Relativity coupled to a scalar field. 
This will lead us to treat the scalar curvature as an independent field and
allow us to propose more general junction conditions than 
(\ref{Jfjunction}) and (\ref{Jfjunctionbis}).

\section{Junction conditions in the Einstein frame\,:
extra degree of freedom and its coupling to matter on the brane}
\label{Einsteinframe}

In order to simplify the notation, we denote quantities in the
Jordan frame with tildes, e.g., $\tilde R$ for the scalar curvature,
and those in the Einstein frame without tildes. Thus all the quantities that
appeared in the previous section should be tilded.

As is well-known \cite{Higgs:1959,Teyssandier:1983zz}, the bulk $f(\tilde R)$ field equations 
for the Jordan frame metric $\tilde g_{AB}$, i.e., $\tilde\sigma_{AB}=0$
with $\tilde\sigma_{AB}$ given by (\ref{sigmadecomp}), 
are equivalent to bulk Einstein equations 
for the ``Einstein frame'' metric,
\begin{equation}
g_{AB}=\tilde g_{AB}\exp\left(\frac{2\phi}{\sqrt{(D-1)(D-2)}}\right)\,,
\end{equation}
with a scalar field minimally coupled to gravity\,:
\begin{equation}
G_{AB}=\partial_A\phi\partial_B\phi
-g_{AB}\left(\frac{1}{2}\partial_C\,\phi\partial^C\phi+W(\phi)\right)\,,
\label{Einscalar}
\end{equation}
where the potential $W(\phi)$ is implicitly defined as a function of
 $\phi$ via
\begin{equation}
W(\tilde R)=\frac{1}{2}
(\tilde R\,f'(\tilde R)-f(\tilde R)){f'(\tilde R)}^{-\frac{D}{D-2}}\,,
\quad
\phi=\sqrt{\frac{D-1}{D-2}}\ln f'(\tilde R)\,.
\end{equation}
Because of the Bianchi identities, the Einstein equations~(\ref{Einscalar})
are consistent only if $\phi$ satisfies the Klein--Gordon equation,
\begin{equation}
\square\phi-\frac{\mathrm dW}{\mathrm d\phi}=0\,.
\end{equation}

Mathematically, this conformal transformation transforms the original, 
fourth-order differential equation (\ref{Jfeq}) into two second-order
differential equations, one for the Einstein frame metric $g_{AB}$,
the other for $\phi$.
Physically, it shows that $f(\tilde R)$ is a ``scalar--tensor'' theory of gravity where $\phi$,
that is, the bulk Jordan frame scalar curvature $\tilde R$, is an extra,
independent, degree of freedom \cite{Damour:1992we}.
What we shall dwell upon in the following is the coupling of this extra 
degree of freedom with matter on the brane.

Again we shall use Gaussian coordinates, that is
\begin{equation}
\mathrm ds^2=\mathrm dz^2+\gamma_{\mu\nu}\mathrm dx^\mu\mathrm dx^\nu\,,
\end{equation}
and consider a class of metrics which are continuous across
 $z=0$, but which allow for discontinuities in their first derivatives. 
We shall also impose the scalar field to be continuous, and allow for 
a discontinuity in its first derivative. Now, $\phi$ is directly related
to the scalar curvature of the bulk Jordan frame metric\,;
one must not however deduce hastily that the condition of continuity of $\phi$
 is equivalent to imposing the continuity of the scalar curvature of 
the Jordan frame metric\,; indeed the relation between
$\phi$ and $\tilde R$ holds {\it in the bulk} only and they may differ,
as we shall see, by a term confined on the brane.

Thus, allowing for discontinuities of the $z$-derivatives of the metric 
and of $\phi$, the right-hand side of the field equations jumps at most.
The $G_{zz}$ and $G_{z\mu}$ components of $G_{AB}$ also jump at most. 
As for the delta-like part of the Einstein tensor $G_{\mu\nu}$, 
it is, see (\ref{Gdecomp})\,:
\begin{equation}
\partial_z(K_{\mu\nu}-K\gamma_{\mu\nu})\equiv\delta(z)D_{\mu\nu}\,.
\end{equation}
Integration across the brane then yields the Israel junction 
conditions \cite{Lanczos}
\begin{equation}
D_{\mu\nu}=[K_{\mu\nu}-K\gamma_{\mu\nu}]^+_-=T_{\mu\nu}\,,
\label{Israel}
\end{equation}
where $T_{\mu\nu}$ is the total energy--momentum tensor of the brane
in the Einstein frame. Later we shall relate it to the
energy--momentum tensor of the brane in the Jordan frame
$\tilde S_{\mu\nu}$.

Since the first derivative of $\phi$ is allowed to be discontinuous,
 $\square\phi$ also exhibits a delta function-like behavior.
From the Bianchi identities,
\begin{equation}
0=\partial_B\phi\left(\square\phi-\frac{\mathrm dW}{\mathrm d\phi}\right)
+D_A(T^A{}_B\delta(z))\,,
\end{equation}
we have
\begin{equation}
\partial_\mu\varphi[\partial_z\phi]^+_-=-\bar D_\nu T^\nu{}_\mu\,,
\label{JCBianchi}
\end{equation}
where $\varphi(x^\mu)=\phi(z=0,x^\mu)$.

Just as in the case of working in the Jordan frame, the last task is
to express the total stress--energy tensor of the brane matter 
in terms of the matter variables.
Since the gravitational variables are $\gamma_{\mu\nu}$ and
$\varphi$ we have to decide how matter couples to those gravitational fields.
 For example, if matter on the brane is taken to be a scalar field $\psi$ 
with potential $V(\psi)$, we may consider as the matter action,
\begin{equation}
S_\mathrm m=-\int\mathrm d^{D-1}x\sqrt{-\gamma}
\left(F_1(\varphi)\frac{1}{2}\gamma^{\mu\nu}\partial_\mu\psi\partial_\nu\psi
+F_2(\varphi)V(\psi)\right)\,,
\label{Smatter}
\end{equation}
where $F_1$ and $F_2$ are two a priori arbitrary functions of $\varphi$.
The associated energy--momentum tensor is 
\begin{equation}
T_{\mu\nu}=F_1(\varphi)\partial_\mu\psi\partial_\nu\psi
-\gamma_{\mu\nu}\left(\frac{1}{2}F_1(\varphi)
\gamma^{\alpha\beta}\partial_\alpha\psi\partial_\beta\psi
+F_2(\varphi)V(\psi)\right)\,.
\end{equation}
The field equation for $\psi$ is
\begin{equation}
\bar D^\nu(F_1(\varphi)\partial_\nu\psi)-F_2(\varphi)V'(\psi)=0\,.
\end{equation}
Thus the divergence of the energy--momentum tensor gives
\begin{equation}
\bar D_\nu T^\nu{}_\mu=-\partial_\mu\varphi
\left(\frac{\mathrm dF_1}{\mathrm d\varphi}\frac{1}{2}
\gamma^{\alpha\beta}\partial_\alpha\psi\partial_\beta\psi
+\frac{\mathrm dF_2}{\mathrm d\varphi}V(\psi)\right)\,.
\end{equation}

Following Einstein's suggestion to Nordstr\"om \cite{Einstein:1913br}, we require
that the source for $\phi$ in (\ref{JCBianchi}) be related to
the trace of the matter energy--momentum tensor,
\begin{equation}
\bar D_\nu T^\nu{}_\mu\propto\partial_\mu\varphi \,T\,.
\label{Tracecond}
\end{equation}
This imposes 
\begin{equation}
F_1(\varphi)=\exp[(D-3)k(\varphi)]\,,
\quad
F_2(\varphi)=\exp[(D-1)k(\varphi)]\,,
\end{equation}
where $k(\varphi)$ still has to be determined. 
Plugging these back into (\ref{Smatter}), we see the meaning of
the condition (\ref{Tracecond}).
Namely, the matter should be minimally coupled to a metric
$\bar\gamma_{\mu\nu}$,
\begin{equation}
S_\mathrm m=S_\mathrm m[\bar\gamma_{\mu\nu}\,;\psi]\,,
\quad
\bar\gamma_{\mu\nu}=\mathrm e^{2k(\varphi)}\gamma_{\mu\nu}\,,
\label{Smcouple}
\end{equation}
where $\psi$ now represents general matter variables
not restricted to a scalar field.

The junction condition (\ref{JCBianchi}) then becomes
\begin{equation}
[\partial_z\phi]^+_-=-\frac{\mathrm dk}{\mathrm d\varphi}\, T\,,
\label{dphijunction}
\end{equation}
with
\begin{equation}
\begin{aligned}
T&=-2\frac{\gamma^{\mu\nu}}{\sqrt{-\gamma}}
\frac{\delta S_\mathrm m[\mathrm e^{2k(\varphi)}\gamma_{\rho\sigma}\,;\psi]}
{\delta\gamma^{\mu\nu}}
\\
&=-\mathrm e^{(D-1)k(\varphi)}
\left(\frac{D-3}{2}\mathrm e^{-2k(\varphi)}\gamma^{\mu\nu}
\partial_\mu\psi\partial_\nu\psi+(D-1)V(\psi)\right)\,,
\end{aligned}
\label{TraceT}
\end{equation}
where the second line is the case when the matter $\psi$ is
a scalar field.
The junction condition (\ref{dphijunction}) is nothing but 
the one used when studying the brane cosmology with
a bulk scalar field \cite{Maeda:2000wrLanglois:2001dy}.

We also note that the matter action can then be rewritten in terms
of the Jordan metric $\tilde\gamma_{\mu\nu}$ as
\begin{equation}
\begin{aligned}
S_\mathrm m
&=S_\mathrm m[\bar\gamma_{\mu\nu}\,;\psi]
=S_m[\mathrm e^{2C(\varphi)}\tilde\gamma_{\mu\nu}\,;\psi]
\\
&=-\int \mathrm d^{D-1}x\sqrt{-\tilde\gamma}\,\mathrm e^{(D-1)C(\varphi)}
\left(\frac{1}{2}\mathrm e^{-2C(\varphi)}
\tilde\gamma^{\mu\nu}\partial_\mu\psi\partial_\nu\psi
+V(\psi)\right)\,,
\end{aligned}
\end{equation}
with
\begin{equation}
 C(\varphi)=k(\varphi)+\frac{\varphi}{\sqrt{(D-1)(D-2)}}\,.
\end{equation}
This is where the fact that we are treating a $f(\tilde R)$ theory in the 
Einstein frame comes into play. Indeed, if we impose 
the matter on the brane 
to be minimally coupled to the Jordan metric, then we must choose
 (a point already known to Einstein \cite{Einstein:1914bu} (see also \citen{Ravndal:2004ym})) $C=0$, that is,
\begin{equation}
k(\varphi)=-\frac{\varphi}{\sqrt{(D-1)(D-2)}}\,.
\label{kchoice}
\end{equation}
This reduces the junction conditions to
\begin{equation}
\begin{aligned}
&[K_{\mu\nu}-K\gamma_{\mu\nu}]^+_-=T_{\mu\nu}\,,
\\
&[\partial_z\phi]^+_-=\frac{T}{\sqrt{(D-1)(D-2)}}\,.
\end{aligned}
\label{Efjunction}
\end{equation}
We shall now translate back to the Jordan frame the generalized 
junction conditions (\ref{Israel}) and (\ref{dphijunction}), and show
that they reduce to those obtained in (\ref{Jfjunctionbis}) 
with (\ref{Jmatter}) only when $k(\varphi)$ 
is imposed to be given by (\ref{kchoice}).

\section{Back to the Jordan frame\,: generalized junction conditions}
\label{Jordanframebis}
In a Gaussian normal coordinate system in which the line element 
reads $\mathrm ds^2=\mathrm dz^2+\gamma_{\mu\nu}\mathrm dx^\mu\mathrm dx^\nu$, the junction conditions 
we have obtained 
in the Einstein frame are
\begin{equation}
\begin{aligned}
&[\phi]^+_-=0\,,
\quad
[\gamma_{\mu\nu}]^+_-=0\,,
\\
&[K_{\mu\nu}-\gamma_{\mu\nu}K]^+_-=T_{\mu\nu}\,,
\quad
[\partial_z\phi]^+_-=-\frac{\mathrm dk}{\mathrm d\varphi}\,T\,,
\end{aligned}
\label{Efsummary}
\end{equation}
where $K_{\mu\nu}=-\frac{1}{2}\partial_z\gamma_{\mu\nu}$
and the second line recalls that the induced metric as well as 
$\phi$ have to be continuous across the brane.
We assume that the matter on the brane couples minimally to
the metric $\bar\gamma_{\mu\nu}=\mathrm e^{2k(\varphi)}\gamma_{\mu\nu}$
 as given by (\ref{Smcouple}).

Let us perform the following transformations
\begin{equation}
\begin{aligned}
&\phi\to\phi=\sqrt{\frac{D-1}{D-2}}\ln f'(\rho)\,,
\\
&W(\phi)\to W(\rho)
=\frac{1}{2}(\rho\,f'(\rho)-f(\rho))
{f'(\rho)}^{-\frac{D}{D-2}}\,,
\\
&g_{AB}\to g_{AB}=\tilde g_{AB}{f'(\rho)}^{\frac{2}{D-2}}\,.
\end{aligned}
\label{phitransform}
\end{equation}
It is a side exercise to show that if $g_{AB}(x^C)$ and $\phi(x^C)$ are 
solution of the bulk field equations (\ref{Einscalar}) then $\rho(x^C)$ is 
the scalar curvature $\tilde R$ of the bulk Jordan metric $\tilde g_{AB}$.
However if one includes the presence of matter on the brane,
there appears a delta function-like singularity in $\tilde R$
while $\rho$ is continuous unless the matter is minimally coupled
on the brane, i.e., unless $C(\varphi)=0$ (modulo a constant),
as we shall see below.

Note that the coordinates are no longer Gaussian
for the Jordan line element 
\begin{equation}
\mathrm d\tilde s^2=\tilde g_{AB}\mathrm dx^A\mathrm dx^B
=f'(\rho)^{-\frac{2}{D-2}}g_{AB}\mathrm dx^A\mathrm dx^B
=f'(\rho)^{-\frac{2}{D-2}}\mathrm dz^2+\tilde\gamma_{\mu\nu}\mathrm dx^\nu\mathrm dx^\nu\,,
\end{equation}
with $\tilde\gamma_{\mu\nu}=f'(\rho)^{-\frac{2}{D-2}}\gamma_{\mu\nu}$ 
the induced metric of the Jordan brane. We introduce
\begin{equation}
\tilde K_{\mu\nu}
=-\frac{1}{2}{f'}^\frac{1}{D-2}\partial_z\tilde\gamma_{\mu\nu}\,,
\quad 
\tilde K=\tilde\gamma^{\mu\nu}\tilde K_{\mu\nu}\,.
\end{equation}

The junction conditions (\ref{Efsummary}) then translate as follows.
As already mentioned, the continuity of
$\phi=\sqrt{\frac{D-1}{D-2}}\ln f'(\rho)$
 translates into the continuity of $\rho$ and 
the continuity of the metric $g_{AB}$ translates into the continuity 
of the Jordan induced metric $\tilde \gamma_{\mu\nu}$\,:
\begin{equation}
[\rho]^+_-=0\,,
\quad
[\tilde \gamma_{\mu\nu}]^+_-=0\,.
\label{rhogamma}
\end{equation}
As for the jumps in $\partial_z\phi$
and the extrinsic curvature they translate into
\begin{equation}
\begin{aligned}
&{f'}^{\frac{1}{D-2}}\,f''[\partial_z\rho]^+_-
=\frac{1}{D-1}\left(1-\sqrt{(D-1)(D-2)}\,
\frac{\mathrm dC}{\mathrm d\varphi}\right)\tilde T\,,
\\
&f'[\tilde K]^+_-
=-\sqrt{\frac{D-1}{D-2}}\,\frac{\mathrm dC}{\mathrm d\varphi}
\,\tilde T\,,
\quad 
f'[\tilde K_{\mu\nu}^*]^+_-
=\tilde T^*_{\mu\nu}\,,
\end{aligned}
\label{general}
\end{equation}
where 
\begin{equation}
\tilde T_{\mu\nu}
=-\frac{2}{\sqrt{-\tilde\gamma}}
\frac{\delta S_\mathrm m[\mathrm e^{2C(\varphi)}\tilde\gamma_{\rho\sigma}\,;\psi]}
{\delta\tilde\gamma^{\mu\nu}}\,,
\label{tildeTmn}
\end{equation}
and a star means taking the traceless part. We note that
because of the assumption that the matter is minimally coupled
to the metric $\bar\gamma_{\mu\nu}$, we have
\begin{equation}
\sqrt{-\tilde\gamma}\,\tilde T=-2\tilde\gamma^{\mu\nu}
\frac{\delta S_m[\mathrm e^{2C(\varphi)}\tilde\gamma_{\rho\sigma}\,;\psi]}
{\delta\tilde\gamma^{\mu\nu}}
=\sqrt{-\gamma}\,T\,.
\end{equation}
If matter on the brane is a scalar field $\psi$, then $S_\mathrm m$
is given by (\ref{dphijunction}) so that
\begin{equation}
\tilde T_{\mu\nu}
=\mathrm e^{(D-1) C(\varphi)}
\left(\mathrm e^{-2C(\varphi)}\partial_\mu\psi\partial_\nu\psi
-\tilde\gamma_{\mu\nu}
\left(\frac{1}{2}\mathrm e^{-2 C(\varphi)}
\tilde\gamma^{\rho\sigma}\partial_\rho\psi\partial_\sigma\psi
+V(\psi)\right)\right)\,.
\end{equation}

Now, generalizing (\ref{Rdecomp}) to a non-Gaussian coordinate,
the scalar curvature $\tilde R$ may be expressed as
\begin{equation}
\tilde R=2N^{-1}\partial_z\tilde K
-\tilde K_{\mu\nu}\tilde K^{\mu\nu}-\tilde K^2
+\tilde{\bar R}\,;
\quad
N=f'(\rho)^{-\frac{1}{D-2}}\,.
\label{tRdecomp}
\end{equation}
Integrating this across $z=0$ and using the
junction conditions (\ref{general}), one finds that
\begin{equation}
\int_{-\epsilon}^\epsilon\tilde R\, Ndz
=2[\tilde K]^+_-
=-2\sqrt{\frac{D-1}{D-2}}\,\frac{\mathrm dC}{\mathrm d\varphi}
\,\frac{\tilde T}{f'}\,.
\end{equation}
From this, we deduce that
\begin{equation}
\tilde R=\rho-2\sqrt{\frac{D-1}{D-2}}\,\frac{\mathrm dC}{\mathrm d\varphi}
\,{f'}^{\frac{1}{D-2}}\,\frac{\tilde T}{f'}\,\delta(z)\,.
\label{Rrhorel}
\end{equation}
Therefore, as anticipated,
the continuity of $\phi$ translates
in the continuity of $\rho$ but not of $\tilde R$\,.

The junction conditions (\ref{rhogamma}) and (\ref{general})
are the central result of this paper.
When the arbitrary function $C$ vanishes,
we have from (\ref{Rrhorel}) that $\tilde R=\rho$ everywhere
including on the brane and we have that 
$\tilde T_{\mu\nu}$ coalesces with $\tilde S_{\mu\nu}$, 
the stress--energy tensor of matter minimally coupled to the brane 
metric introduced in (\ref{JfKjunction}).
The junction conditions thus reduce to those obtained in
Sec.~\ref{Jordanframe}.

When $C$ is a non-trivial function,
they generalize them to the case when matter on the 
brane is coupled not only to the brane metric $\tilde\gamma_{\mu\nu}$ but
to the extra degree of freedom of $f(\tilde R)$ gravity which, 
in the Jordan frame, is the quantity $\rho$, equal everywhere but on
the brane to the scalar curvature $\tilde R$.

We just note here, to conclude, that, if we choose 
$C(\varphi)=\varphi/\sqrt{(D-1)(D-2)}$, that is $k(\varphi)=0$, 
then the junction conditions in the Jordan frame
closely resemble the standard Israel junction conditions
\begin{equation}
f'[\tilde K_{\mu\nu}-\tilde K\tilde\gamma_{\mu\nu}]^+_-
=\tilde T_{\mu\nu}\,.
\end{equation}
However the coupling of matter 
to the brane metric is not minimal, as given by (\ref{tildeTmn}).

\section{Conclusion}
We thoroughly investigated the junction conditions in $f(R)$
theories of gravity. We found that in a pure
 $f(R)$ theory in which matter on the brane couples minimally
to the metric, the bulk scalar curvature $R$
must be continuous across the brane, which is in marked contrast
with the case of Einstein gravity. Then taking advantage of
the conformal equivalence of $f(R)$ theories with
Einstein gravity with a scalar field, we clarified the
importance of identifying the scalar curvature $R$
as an extra degree of freedom and the specific form of the 
coupling of matter to this extra gravitational degree of freedom
on the brane.

Then as a bonus of working in the Einstein frame, 
we presented a natural generalization 
of the coupling of matter to gravity. In the original frame,
this leads to a non-trivial coupling of the matter on the brane
to the extra degree of freedom, which allows a delta function-like
behavior of the scalar curvature. This suggests a new class
of braneworld models whose solutions may have a smooth limit
in the Einstein limit $f(R)\to R$.

It is known that, in the bulk,
an $f(R)$ theory may be rewritten as a Brans--Dicke theory
with $\omega=0$ but with a potential \cite{Chiba:2003ir}. 
If we use this equivalence, the generalization mentioned above
may be regarded as a non-trivial coupling of the Brans--Dicke
scalar to the matter on the brane.

We leave to an accompanying paper the derivation of the junction 
conditions (\ref{general}) via a first order description of the $f(R)$
action as well as an analysis of the braneworld models 
that they may lead to.

\section*{Acknowledgements}
This work was supported in part by JSPS Grant-in-Aid for
Scientific Research (B) No.~17340075, (A) No.~18204024, 
and by JSPS Grant-in-Aid for Creative Scientific Research No.~19GS0219.
YS was also supported by Grant-in-Aid for JSPS Fellows No.~7852.
We are grateful to  Valeri Frolov, Jaume Garriga and Sergei Odintsov for
fruitful discussions.
ND thanks the Yukawa Institute for Theoretical Physics at Kyoto 
University, where this work was partially done during the scientific
program on ``Gravity and Cosmology 2007'' and
the workshop YITP-W-07-10 on ``String Phenomenology and Cosmology.'' 
MS and ND also acknowledge financial support 
from the CNRS--JSPS Exchange Programme ``Sakura.''

\end{document}